\documentclass[10pt,conference]{IEEEtran}
\usepackage{cite}
\usepackage{amsmath,amssymb,amsfonts}
\usepackage{algorithmic}
\usepackage{graphicx}
\usepackage{textcomp}
\usepackage{xspace}

\usepackage[bookmarks=false]{hyperref}
\hypersetup{
    colorlinks=true,
    allcolors=blue,
}
\usepackage{cleveref}

\usepackage[ff,sets,adversary,keys,primitives,operators,lambda,logic,notions]{cryptocode}
\usepackage{boxedminipage}

\def\systemnameRaw{VDKMS}
\def\systemname{\systemnameRaw\xspace}

\newcommand{\registrationCredential}{VIN Credential\xspace}

\newcommand{\protocol}[3][\columnwidth]{
	\begin{boxedminipage}[t]{#1}
		\begin{center}
		\scriptsize{\textbf{#2}}
		\end{center}
		\vspace{-\baselineskip}
		\procedure[mode=text, linenumbering, codesize=\scriptsize]{}{			
			#3
		}
	\end{boxedminipage}
}

\newcommand{\ekdparty}{{\mathcal{P}}}

\newcommand{\onrecv}{{\bf On receive}\xspace}

\def\BibTeX{{\rm B\kern-.05em{\sc i\kern-.025em b}\kern-.08em
    T\kern-.1667em\lower.7ex\hbox{E}\kern-.125emX}}

\begin{document}

\title{\systemname{}: Vehicular Decentralized Key Management System for Cellular Vehicular-to-Everything Networks, A Blockchain-Based Approach
}

\author{\IEEEauthorblockN{Wei Yao\IEEEauthorrefmark{1}, Yuhong Liu\IEEEauthorrefmark{2}, Fadi P. Deek\IEEEauthorrefmark{1} and Guiling Wang\IEEEauthorrefmark{1}} 
\IEEEauthorblockA{\IEEEauthorrefmark{1}Ying Wu College of Computing\\
New Jersey Institute of Technology, Newark, NJ 07102\\
Email: wy95, fadi.deek and gwang @njit.edu}
\IEEEauthorblockA{\IEEEauthorrefmark{2}Computer Engineering Department\\
Santa Clara University, Santa Clara, CA 95053\\
Email: yhliu@scu.edu}}

\maketitle

\begin{abstract}
The rapid development of intelligent transportation systems and connected vehicles has highlighted the need for secure and efficient key management systems (KMS). In this paper, we introduce VDKMS (Vehicular Decentralized Key Management System), a novel Decentralized Key Management System designed specifically as an infrastructure for Cellular Vehicular-to-Everything (V2X) networks, utilizing a blockchain-based approach. The proposed VDKMS addresses the challenges of secure communication, privacy preservation, and efficient key management in V2X scenarios. It integrates blockchain technology, Self-Sovereign Identity (SSI) principles, and Decentralized Identifiers (DIDs) to enable secure and trustworthy V2X applications among vehicles, infrastructures, and networks. We first provide a comprehensive overview of the system architecture, components, protocols, and workflows, covering aspects such as provisioning, registration, verification, and authorization. We then present a detailed performance evaluation, discussing the security properties and compatibility of the proposed solution, as well as a security analysis. Finally, we present potential applications in the vehicular ecosystem that can leverage the advantages of our approach.
\end{abstract}


\section{Introduction} 
With the rapid growth of connected and autonomous vehicles, vehicular networks are becoming increasingly important, and Key Management Systems (KMS) play a critical role in ensuring the security and integrity of these networks. Public Key Infrastructure (PKI) is a widely adopted KMS that provides secure communication, authentication, and data encryption among vehicles. However, despite its significance, the current implementation of PKI in vehicular networks faces several issues that undermine its effectiveness.
The primary challenges faced by PKI in vehicular networks include: (1) Compatibility issues—The limitations of Vehicular Ad hoc Networks (VANETs) create compatibility challenges, such as supporting computer-network based communication protocols that are essential for secure data exchange among vehicles. (2) Single points of failure—The centralized nature of PKI introduces risk for faults, however small, that can compromise the entire system’s security and stability if it is attacked or simply malfunctions. (3) Complex verification process—The trust model in PKI leads to an intricate verification method, which can be inefficient and resource-intensive, especially for time-sensitive vehicular communication. (4) Privacy preservation—Ensuring the correctness of the verification process while also hiding sensitive privacy information is a major concern. Achieving a balance between privacy preservation and secure communication is vital for maintaining user trust and the overall functionality of vehicular networks. Given these challenges, it is essential to explore alternative solutions, such as Decentralized Key Management Systems (DKMS), which can address these issues and provide a more secure, efficient, and resilient framework for vehicular communication. However, implementing DKMS in existing VANETs and their communication systems presents its own set of limitations and complications that need to be addressed in order to fully harness the potential of decentralized vehicular networks. Thus, we propose a novel Vehicular Decentralized Key Management System (VDKMS) for Cellular Vehicular-to-Everything (V2X) networks, which leverages the advancements in Self-Sovereign Identity (SSI) and Decentralized Identifier (DID) to create a secure, scalable, and efficient key management schema, specifically tailored to overcome the limitations of implementing DKMS in existing VANETs and their communication systems.

The main contributions of this paper are as follows: (1) A comprehensive design of VDKMS that is specifically tailored to address the unique challenges and requirements of vehicular networks, as well as the limitations of implementing DKMS in existing VANETs. (2) A proposal for a novel integration of SSI and DID within the VDKMS to enhance security, privacy, and trust in the system, while also ensuring the system’s scalability and adaptability to future advancements in vehicular networks. (3) An extensive analysis of the proposed VDKMS, including its performance and security, as well as its compatibility and interoperability with existing vehicular communication standards and protocols. (4) We showcase potential applications within the vehicular ecosystem that can effectively utilize the benefits provided by our VDKMS.

The remainder of this paper is organized as follows: Section \ref{sec:RL} presents related work on decentralized key management systems, vehicular communication, and SSI technology. Section \ref{sec:system} provides an overview of the proposed Decentralized Cellular Vehicular-to-Everything Key Management System and its components. Section \ref{sec:protocols} discusses the integration of SSI and DID technology within the DKMS. Section \ref{sec:evaluation} presents the performance, security, and privacy analysis of the proposed system. Section \ref{sec:applications} discusses the type of applications that can benefit from our approach. Finally, Section \ref{sec:conclusion} concludes the paper and discusses potential directions for future research.
\section{Related Work} \label{sec:RL}
The need for a secure and efficient key management in vehicular networks has been widely recognized in the literature. Several approaches have been proposed to address the challenges of traditional Key Management Systems (KMS) in this domain, and Decentralized Key Management Systems (DKMS) have emerged as a promising solution. In this section, we review the most relevant work related to DKMS and highlight their limitations, categorizing them into three groups: (1) KMS in vehicular networks: A significant amount of research has been conducted in this domain, such as \cite{asghar2018scalable,hesham2011dynamic,wasef2010complementing}. These approaches typically rely on a centralized authority to manage cryptographic keys and secure communication between vehicles. However, centralized KMS suffer from several disadvantages, such as single points of failure, limited scalability, and potential privacy breaches. These issues have motivated researchers to explore alternative approaches. (2) DKMS in vehicular networks without leveraging DID, SSI, and blockchain: The use of DKMS to address the limitations of centralized KMS in vehicular networks has been proposed in \cite{li2018creditcoin,lu2019blockchain,yao2019bla,ma2020efficient}. These approaches leverage blockchain technology, distributed ledgers, and other decentralized architectures to provide more secure, scalable, and efficient key management systems. However, most existing DKMS proposals for vehicular networks have not considered the integration of Self-Sovereign Identity (SSI) and Decentralized Identifier (DID) technology, which can further enhance the security, privacy, and trust in these systems. (3) Use of DID, SSI, and blockchain for personal identity management: The integration of DIDs and blockchain technology for identity management has been explored in a number of studies such as CanDID \cite{maram2021candid}. This body of work primarily focuses on the domain of personal identity management, featuring a unique set of use-cases and challenges distinct from vehicular networks. In contrast, our work aligns more specifically with the context of cellular vehicular networks, taking into account the particular needs and constraints of this field.


\section{System Architecture and Components} \label{sec:system}
\begin{figure}[htbp]
    \centering
    \includegraphics[width=0.95\linewidth]{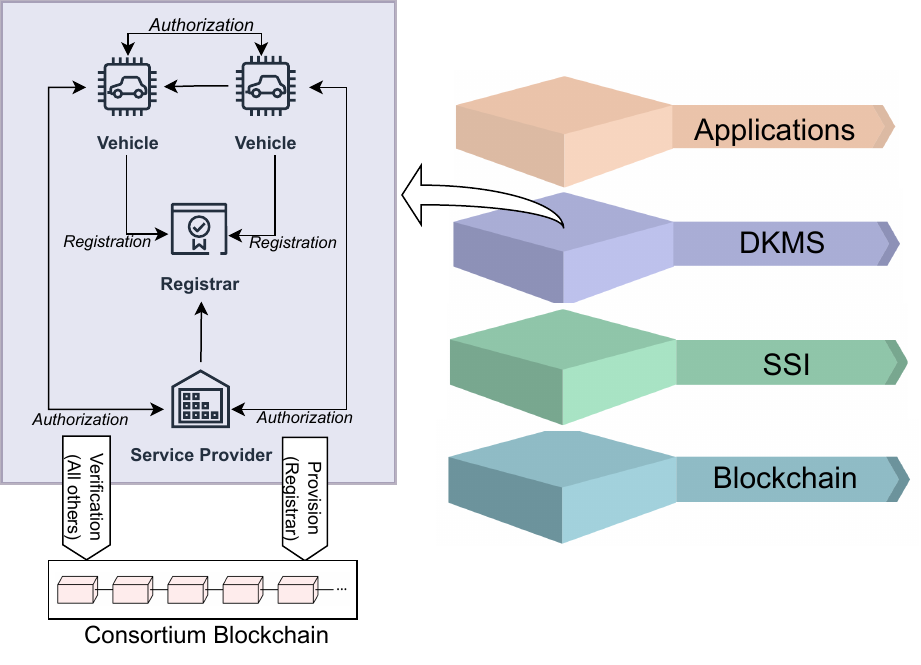}
    \caption{VDKMS Architecture}
    \label{fig:archi}
\end{figure}

Cellular Vehicular-to-Everything (C-V2X) networks are communication systems designed to enable a wide range of applications and services for connected vehicles. These networks facilitate seamless and secure data exchange between vehicles (V2V), vehicles and infrastructures (V2I), vehicles and networks (V2N), and vehicles and pedestrians (V2P). Key participants in C-V2X networks include vehicles, infrastructure components such as roadside units (RSUs), communication service providers, network operators, and registrars. \systemname{} plays a crucial role in C-V2X networks by providing a secure, scalable, and efficient key management framework that enables trustworthy and reliable communication among network participants. It ensures the privacy and integrity of exchanged data while also maintaining the adaptability of the system to future advancements in vehicular communication. In this section, we present the architecture and components of our proposed \systemname{}, shown in Figure \ref{fig:archi}.

\subsection{Participants}
The participants includes: 
(1) Registrars: A category of entities that serve as Certification Authority (CA) with the permission to issue vehicle credentials and write transactions to the blockchain. A registrar can be a high-profile government-owned institution responsible for running and maintaining the blockchain infrastructure. For instance, the New York Department of Transportation.
(2) Vehicles: Participants in the C-V2X network that require \systemname{} as the core mechanism to ensure the privacy, integrity, and authenticity of exchanged data, addressing various use scenarios in the C-V2X network.
(3) Service Providers: Entities representing various aspects of the C-V2X network, such as roadside infrastructure, traffic management systems, or other services that interact with vehicles. Service providers also require \systemname{} to ensure the privacy, integrity, and authenticity of exchanged data. In certain cases, service providers may issue credentials to vehicles, authorizing them to access specific services.

\subsection{Data Models}
Decentralized Identities (DIDs): A unique, persistent, and cryptographically verifiable identifier for vehicles, infrastructure components, and other network participants. DIDs are registered on the blockchain-based infrastructure, ensuring their immutability and verifiability. They are designed and implemented following the World Wide Web Consortium's (W3C) Decentralized Identifiers (DIDs) specification \cite{did_specification}.

Verifiable Credentials (VCs): Digital representations of credentials that include information about the subject, identification data, and claim(s) describing the subject. VCs are issued by registrars or service providers, cryptographically signed, and the verification data is stored on a blockchain, making them tamper-resistant and verifiable. VCs are in compliance with the W3C's Verifiable Credentials specification \cite{verifiable_credential_specification}. There are three types of credentials in C-V2X networks: (1) Vehicle Credentials, an issued proof of qualification or identity for a vehicle, including information such as vehicle identification number (VIN), make, model, and owner details. The VIN is closely linked with the vehicle's DID in the Registration Credentials, so each vehicle can have only one Registration Credentials. (2) Service Provider Credentials, an issued proof of qualification for a service provider by a registrar, enabling the service provider to issue regular credentials to vehicles. (3) Regular Credentials, a proof of authorization for a vehicle issued by a service provider.

\subsection{Components}
Our \systemname{} consists of the following components: (1) Blockchain-based Infrastructure: The blockchain-based infrastructure serves as the foundation for our proposed \systemname{}. It provides a decentralized and tamper-resistant platform for managing cryptographic keys and maintaining an immutable record of key-related transactions. The blockchain infrastructure is built on a consensus mechanism that ensures the integrity and consistency of stored data. In our system, we adopt a consortium blockchain \cite{yao2021survey}, bridging the gap between the transparency of public blockchains and the centralized control of private blockchains. In this setup, a select group of nodes, which could include certain infrastructure components or trusted entities within the vehicular network, are entrusted with the responsibility of validating transactions and appending new blocks. This configuration offers multiple benefits to our system. Firstly, it significantly bolsters the transaction throughput while diminishing the latency, thereby mitigating the scalability concerns frequently tied to public blockchains. In addition, by permitting only reliable nodes to engage in consensus, it bolsters the security robustness and the resilience against Sybil attacks.
(2) Self-Sovereign Identity and Decentralized Identifiers: Our \systemname{} leverages Self-Sovereign Identity (SSI) and Decentralized Identifier (DID) technology to provide secure, private, and user-centric identity management for vehicles, users, and other entities within the vehicular network. SSI allows individuals and entities to own and control their digital identities without relying on a centralized authority. DIDs are globally unique identifiers that can be generated, resolved, and managed in a decentralized manner. By integrating SSI and DID in our \systemname{}, we can enhance the security, privacy, and trust in the system while ensuring its scalability and adaptability to future advancements in vehicular communication.
(3) V2X Cryptographic Key Management, is responsible for managing the cryptographic keys used for secure V2X communication. It comprises two main sub-components: (a) Key Generation and Distribution: Our \systemname{} facilitates the secure generation and distribution of cryptographic keys for V2X communication. The key generation process involves creating public-private key pairs for vehicles, users, and other entities within the vehicular network. The public keys are stored on the blockchain, while the private keys are securely stored by the entities themselves. Key distribution is performed using secure communication channels and cryptographic protocols, ensuring that only authorized entities can access the required keys. (b) Key Revocation and Renewal: The \systemname{} also provides mechanisms for key revocation and renewal. Key revocation is necessary when a vehicle or user leaves the network or in case of a security breach. Our \systemname{} leverages the blockchain infrastructure to implement a secure and efficient key revocation process. Key renewal is performed periodically or on-demand to maintain the freshness and security of the cryptographic keys. The key renewal process is also automated, ensuring a seamless and secure update of keys within the vehicular network.
\section{Protocols and Workflow} \label{sec:protocols}
In this section, we describe the protocols and workflow involved in our proposed VDKMS. We outline the protocols, focusing on registrar and vehicle provisions, vehicle registration as identity management, key establishment, and secure communication, as well as credentials verification. Our VDKMS is designed to facilitate secure and efficient communication and key management within the vehicular network. The following protocols shown in Figure \labelcref{fig:provision,fig:registration,fig:verification,fig:authorization} are proposed to achieve these objectives:

\subsection{Provisions}
\begin{figure}
\protocol{Provision} 
{\\
\underline{\bf Registrar $\mathcal{R}$:} \\
\t Generates key pair through ($\mathsf{pk}^\mathcal{R}$, $\mathsf{sk}^\mathcal{R}$, $\mathsf{DID}^\mathcal{R}$) $\gets$ $\mathsf{KeyGen}$() \\
\t $\mathsf{RCSchema}$ $\gets$ $\mathsf{ScheGen} (\{attr\}, \mathsf{sk}^\mathcal{R})$\\
\t $\mathsf{RCDef}$ $\gets$ $\mathsf{DefGen} (\mathsf{RCSchema}, \mathsf{pk}^\mathcal{R})$ \\
\underline{\bf Vehicle $\mathcal{V}$:} \\
\t Generates key pair through ($\mathsf{pk}^\mathcal{V}$, $\mathsf{sk}^\mathcal{V}$, $\mathsf{DID}^\mathcal{V}$) $\gets$ $\mathsf{KeyGen}$() \\
\underline{\bf Service Provider $\mathcal{SP}$:} \\
\t Generates key pair through ($\mathsf{pk}^\mathcal{SP}$, $\mathsf{sk}^\mathcal{SP}$, $\mathsf{DID}^\mathcal{SP}$) $\gets$ $\mathsf{KeyGen}$() 
}
\caption{Provision}
\label{fig:provision}
\end{figure}

(1) Registrar Provision: This workflow describes the process of initializing a registrar capable of registering a Decentralized Identifier (DID) and writing it to the blockchain ledger. The workflow starts with the registrar generating an asymmetric keypair from its digital wallet. The protocol allows the registrar to create the asymmetric keypair and its DID, and the DID can be securely stored. The encryption process involves encrypting the storage with a wallet encryption key. After this step, the registrar holds the key pair ($\mathsf{pk}^\mathcal{R}$, $\mathsf{sk}^\mathcal{R}$) and $\mathsf{DID}^\mathcal{R}$.
The registrar creates the schema $\mathsf{RCSchema}$ of \registrationCredential which contains attributes $\{\mathsf{attr_i}\}$ and registers the schema on the ledger through a blockchain transaction. 
Based on the schema of \registrationCredential, the registrar creates and registers a credentials definition $\mathsf{DefGen}$ on the blockchain. The registrar's identity information and cryptographic attributes are bound to the credentials schema and used for verification purposes. 
(2) Vehicles Provision:
The vehicle initialization process begins with the creation of a key pair and a DID for the vehicle. After the vehicle provision, a vehicle $\mathcal{V}$ holds key pairs ($\mathsf{pk}^\mathcal{V}$, $\mathsf{sk}^\mathcal{V}$) and DID $\mathsf{DID}^\mathcal{V}$. 
(3) Service Providers Provision:
The service provider $\mathcal{SP}$ also holds key pair ($\mathsf{pk}^\mathcal{SP}$, $\mathsf{sk}^\mathcal{SP}$) and DID $\mathsf{DID}^\mathcal{SP}$ after the provision. This allows them to securely interact with all participants in the C-V2X network.

\subsection{Registration as Identity Management}
\begin{figure}
\protocol{Registration} 
{\\
\underline{\bf Registrar $\mathcal{R}$:} \\
\onrecv ($\mathsf{DID}^\mathcal{V}$, $\{\mathsf{attr}\}$): \\
\t \pcif $\mathsf{VIN}$ $\in$ $\{\mathsf{attr}\}$ and $\{\mathsf{attr}\}$ is $\mathsf{valid}$: $\sigma$ $\gets$ $\mathsf{Sign_{\mathsf{sk}^\mathcal{R}}}$(${\mathsf{\{claim\}}}$, $\mathsf{pk}^\mathcal{V}$) \\
\t $\mathsf{VCred}$ $\gets$ $\mathsf{CredGen}$ ($\mathsf{Schema}$, $\sigma$, ${\mathsf{\{claim\}}}$, $\mathsf{pk}^\mathcal{V}$) \\
\t $\mathsf{txn}$ $\gets$ ($\mathsf{VCred}$, $\mathsf{DID}^\mathcal{V}$) \\
\underline{\bf Vehicle $\mathcal{V}$:} \\
\t Send request ($\mathsf{DID}^\mathcal{V}$, $\{\mathsf{attr}\}$), while $\mathsf{VIN}$ $\in$ $\{\mathsf{attr}\}$\\
\t \pcif $\mathsf{VIN}$ $\in$ $\{\mathsf{attr}\}$ and $\{\mathsf{attr}\}$ is $\mathsf{valid}$: Receive $\mathsf{VCred}$
}
\caption{Registration}
\label{fig:registration}
\end{figure}

Vehicle registration is a critical step for managing the identity of a vehicle in the \systemname{}. The vehicle sends a registration request containing its DID and other necessary information as attributes to the registrar. The registrar checks whether the VIN is included and whether all information as the value of the vehicle's attributes are valid. If everything is verified, the registrar will confirm all the values of attributes as ${\mathsf{claims}}$ and sign ${\mathsf{claims}}$.
Then, the registrar issues a \registrationCredential to the vehicle, associating the vehicle's DID.
Once the Vehicle Credentials is issued, the registrar writes the DID, Vehicle Credentials, and associated public keys through a blockchain transaction $\mathsf{txn}$. This ensures the immutability and verifiability of the vehicle's identity and credentials, allowing other network participants to trust and authenticate the vehicle during communication.
Upon completing the registration, the vehicle receives a confirmation message along with the signed credentials. This credentials serves as proof of the vehicle's identity and registration in the vehicular network. To enhance communication efficiency, the registrar employs elliptic curves to create signatures, effectively reducing their size \cite{camenisch2016anonymous}.

The service provider registration process shares similarities with the vehicle registration process. However, the key difference lies in the schema for service provider registration, which can be more customized since it is not based on a master ID like the VIN in vehicle registration. The creation of a schema for service provider registration is beyond the scope of this discussion, but it can be tailored to the specific requirements of the service providers in the network, allowing for greater flexibility and adaptability.

\subsection{Credentials Verification}
\begin{figure}
\protocol{Credentials Verification} 
{\\
\textbf{Input}: Vehicle $\mathcal{V}$ inputs $\mathsf{cred}$ and $\mathsf{sk}^\mathcal{V}$. Prover $\mathcal{P}$ inputs a challenge $\mathsf{c}_\mathsf{VIN}$. \\
\textbf{Output}: $\ekdparty$ outputs $\mathsf{success}$ or $\mathsf{fail}$.\\
\underline{\bf Vehicle $\mathcal{V}$:} \\
\t $\mathsf{\Pi}$ $\gets$ $\mathsf{ProofGen}$ ($\mathsf{VCred}$, $\mathsf{c}_\mathsf{VIN}$), $\sigma$ $\gets$ $\mathsf{Sign}_{\mathsf{sk}^\mathcal{V}}(\mathsf{\Pi})$  \\
\t Send ($\mathsf{\Pi}$, $\sigma$, $\mathsf{DID}^\mathcal{V}$, $\mathsf{DID}^\mathcal{R}$) to $\mathcal{P}$ \\ 
\underline{\bf Prover $\ekdparty$:} \\
\t Check if $\mathsf{SigVerify}$ ($\mathsf{\Pi}$, $\sigma$, $\mathsf{DID}^\mathcal{V}.\mathsf{pk}^\mathcal{V}$) = 1 $\land$ $\mathsf{\Pi}$.$\mathsf{issuer}$ = $\mathsf{DID}^\mathcal{R}$ $\land$ \newline $\mathsf{\Pi}$.$\mathsf{subject}$ = $\mathsf{DID}^\mathcal{V}$ $\land$
$\mathsf{\Pi}$.$\mathsf{claim}$ contains $\mathsf{VIN}$ 
} 
\caption{Credentials Verification}
\label{fig:verification}
\end{figure}

Credentials verification is a crucial process in ensuring the validity of a vehicle's identity and its associated credentials. In \systemname{}, when a vehicle requires authorization from a service provider, it sends a request to the service provider. In response, the service provider sends a proof request for a specific attribute of the VIN, denoted as a challenge $\mathsf{c}_\mathsf{VIN}$. Upon receiving the challenge and possessing a \registrationCredential, the vehicle generates a presentation from the original \registrationCredential that includes the value of the VIN as the requested attribute. The presentation contains a proof denoted as $\mathsf{\Pi}$, used for verification. The vehicle digitally signs it, obtaining the signature $\sigma$, which claims ownership of the proof. The vehicle then sends the presentation, which contains $\mathsf{\Pi}$, $\sigma$, $\mathsf{DID}^\mathcal{V}$, and $\mathsf{DID}^\mathcal{R}$, to the service provider.
The service provider first checks whether a trustworthy registrar has issued the credentials in the presentation. It then verifies if the presented attributes in the proof satisfies the request. The recipient of the credentials confirms its authenticity by checking the signature against the public key of the issuing registrar. If the signature is valid, the recipient accepts the vehicle's credentials and proceeds with further communication. This verification process ensures that only vehicles with valid credentials can participate in secure communication within the V2X network.

\subsection{Authorization}
\begin{figure}
\protocol{Authorization} 
{\\
\textbf{Preconditions}: The Service Provider has a well-known DID as $\mathsf{DID}^\mathcal{SP}$ \\
\underline{\bf Vehicle $\mathcal{V}$:} \\
\t Generates $\mathsf{DID}_{\mathcal{SP}}$ and send request ($\mathsf{DID}^{\mathcal{V}}$, $\mathsf{DID}_{\mathcal{SP}}$)\\
\onrecv (${\mathsf{DID}_{\mathcal{V}}}^{\mathcal{SP}}$): \\
\t $\mathsf{txn}_\mathsf{microledger}$ $\gets$ (${\mathsf{DID}}^{\mathcal{V}}$, ${\mathsf{DID}_{\mathcal{V}}}^{\mathcal{SP}}$) \\
\underline{\bf Service Provider $\mathcal{SP}$:} \\
\onrecv ($\mathsf{DID}^{\mathcal{V}}$, $\mathsf{DID}_{\mathcal{SP}}$): \\ 
\t If \textbf{Verification} is $\mathsf{valid}$: \\ 
\t \t Generates ${\mathsf{DID}_{\mathcal{V}}}^{\mathcal{SP}}$ only used as private DID used between $\mathcal{V}$ and $\mathcal{SP}$ \newline and key pair through (${\mathsf{pk}_{\mathcal{V}}}^{\mathcal{SP}}$, ${\mathsf{sk}_{\mathcal{V}}}^{\mathcal{SP}}$)\\
\t \t $\mathsf{txn}_\mathsf{microledger}$ $\gets$ (${\mathsf{DID}_{\mathcal{V}}}^{\mathcal{SP}}$, ${\mathsf{DID}}^{\mathcal{V}}$) \\
\t \t Sends (${\mathsf{DID}_{\mathcal{V}}}^{\mathcal{SP}}$) to the sender vehicle $\mathcal{V}$ 
}
\caption{Authorization}
\label{fig:authorization}
\end{figure}
In order for a service provider to provide services to a vehicle, it must first verify that the vehicle has a valid vehicle credentials. The verification process is described in the previous subsection, Credentials Verification. After successful verification, the service provider and the vehicle generate peer DIDs, which are private DIDs that serve only for communication between the two parties. These private DIDs are stored in a local storage known as a microledger, which is a similar data structure as a blockchain ledger.
Once the vehicle has been authorized, it gains access to a private communication channel to use the services provided by the service provider. This private channel ensures a secure and efficient connection between the vehicle and the service provider. As a result, no additional verifications are needed for subsequent interactions between the vehicle and the service provider. This authorization process establishes trust between the two parties, enabling secure and seamless access to services within the V2X network.

\section{Evaluation and Security Analysis} \label{sec:evaluation}
In this section, we discuss the methodology used for evaluating the performance of our proposed \systemname{} and present the results of our evaluation. Additionally, we provide a security analysis of the system.

\subsection{Experimental Evaluation}
\begin{figure}[htbp]
    \centering
\includegraphics[width=\linewidth]{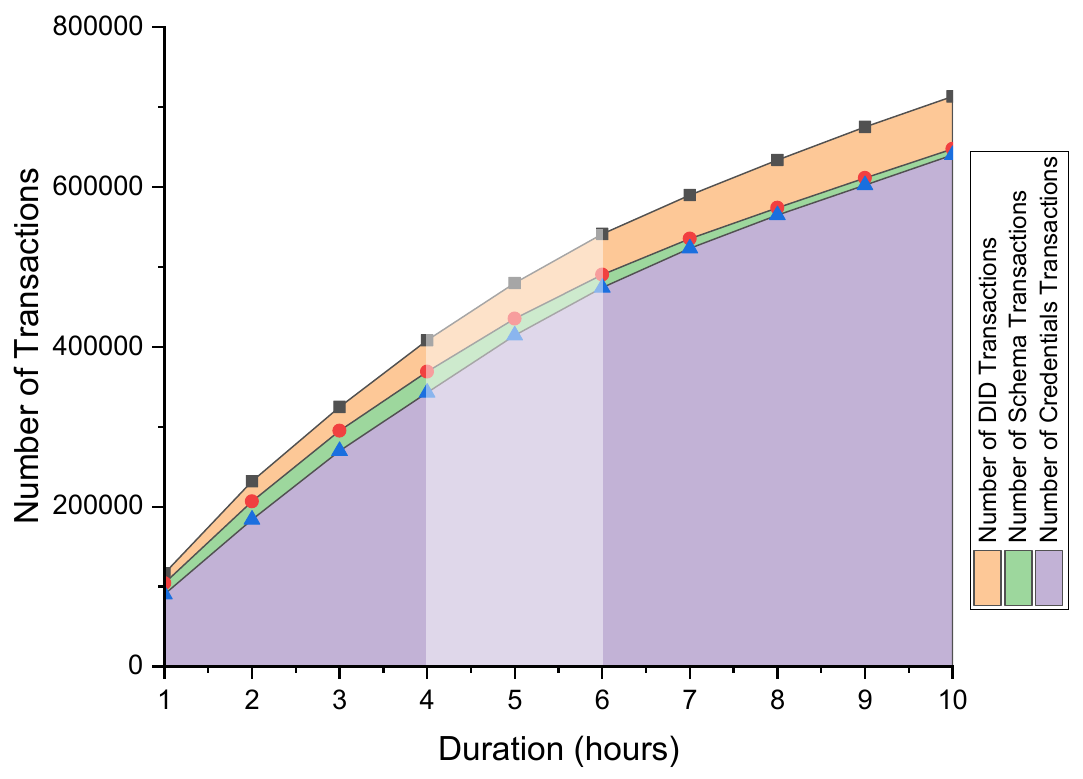}
    \caption{Performance while interruption happens.}
    \label{fig:vstat}
\end{figure}

To evaluate the performance and scalability of our proposed \systemname{} approach, we deployed a blockchain ledger on an Amazon Web Services (AWS) EC2 t2.xlarge instance, with up to 3.3 GHz Intel Xeon Scalable processor and 16 GB memory, employing the Plenum Byzantine Fault Tolerant Protocol \cite{8456064}. Plenum is an implementation of the RBFT protocol. In contrast to the RBFT protocol's use of message authentication codes (MACs), Plenum adopts elliptic curve cryptography for signing communications. The enhanced security offered by elliptic curve cryptography justifies the computational overhead of signature verification. The blockchain ledger was deployed on seven nodes to simulate a real-world scenario. Our experimental setup encompasses three types of transactions, corresponding to the provision and registration phases mentioned earlier. The sizes of these transactions vary due to differences in their contents. We conducted our experiments for a duration of 10 hours, with the blockchain nodes experiencing an interrupted status between the 4th and 6th hour of the experiment. During this time, only four out of the seven nodes were running, but no significant interruptions were observed. The results shown in Figure \ref{fig:vstat} demonstrate that our proposed \systemname{} is capable of processing a large number of transactions efficiently, even under interrupted conditions. This indicates the resilience and robustness of the system in handling real-world vehicular communication scenarios. The scalability of the system is also evident from the increasing number of transactions processed over time, showcasing its ability to support the growing demands of the vehicular ecosystem.

Our DKMS is compatible with existing vehicular communication standards and protocols, such as Dedicated Short-Range Communications (DSRC), IEEE 802.11p, HTTP, and XMPP, ensuring seamless integration with current vehicular networks. The interoperability of the system is also noteworthy, as it can work with different blockchain platforms including Ethereum and Hyperledger Fabric.

\subsection{Security Analysis}
In this section, we examine various security aspects of the system, focusing on confidentiality, integrity, and authentication. \systemname{} achieves these objectives by leveraging DID-based communication protocols, such as the DIDComm specification \cite{curren2023didcomm}.
\textit{Confidentiality}: The DKMS ensures the confidentiality of communications between network entities by employing end-to-end encryption protocols. This makes certain that even vehicles or users with valid IDs cannot eavesdrop on the communications between two separate entities within the system.
Strong cryptographic algorithms are employed for message encryption to guarantee that the content remains confidential and accessible only to the intended recipients.  
These secure communication channels complement the existing authentication mechanisms, such as DIDs and SSIs, which ensure that only authorized entities can attempt to communicate within the network.
\textit{Integrity}: The DKMS guarantees the integrity of data exchanged in the vehicular network by employing robust cryptographic techniques, such as digital signatures and hash functions. This ensures that any unauthorized modifications to data can be detected and prevented. 
\textit{Authentication}: The DKMS provides strong authentication mechanisms for vehicles, users, and other entities in the network, ensuring that only authorized parties can access and use the system. The use of DIDs and SSI mechanisms further strengthens the authentication process. 

The mechanisms in the DKMS that ensure confidentiality, integrity, and authentication also contribute to its resistance against eavesdropping and man-in-the-middle attacks. Moreover, replay attacks are neutralized by incorporating a unique timestamp and nonce in each transmitted message, which confines its validity to a specific time frame and eliminates the possibility of unauthorized future reuse. 
The decentralized nature of the system, combined with the immutable characteristics of blockchain technology, add additional layers of resistance against these and other potential attacks. For example, the immutability of blockchain records makes tampering extremely difficult, if not impossible. Each transaction is verified by multiple nodes in the network, ensuring integrity and making large-scale fraud unlikely.
The distributed ID management system further fortifies the network's defenses against false information injection attacks by enabling the identification and penalization of dishonest vehicles that disseminate false information.

Overall, our security analysis demonstrates that the proposed \systemname{} provides a high level of security for vehicular networks, ensuring the confidentiality, integrity, and authentication of data and communications. The system's resistance to various security attacks further enhances its reliability and trustworthiness in the field of vehicular communications.
\section{Applications} \label{sec:applications}
The proposed blockchain-based \systemname{} approach provides a robust and secure foundation for various applications in the vehicular ecosystem. The following potential applications can leverage the advantages of our approach: 
(1) Vehicle-to-Vehicle (V2V) Communication, critical for connected and autonomous vehicles, our DID-based verification significantly improves the security and reliability of V2V communications. Only vehicles with verified credentials can participate, offering protection against attacks like spoofing and replay attacks.
(2) Vehicle-to-Infrastructure (V2I) Services, ensures secure communications between vehicles and roadside facilities like traffic lights and toll booths. Leveraging cellular networks for connectivity, our system utilizes DID and SSI technology for secure and private authentication, facilitating real-time data exchange. Moreover, it keeps a transparent and immutable record of transactions, aiding in dispute resolution.
(3) Extended Vehicular Ecosystem: Our approach is not limited to V2V and V2I. It can be applied to other services in the vehicular environment, such as vehicle financing, maintenance, and emissions monitoring. The decentralized platform adds trust and transparency, mitigating inefficiencies and potential fraud in traditional systems.
Moreover, our approach can be extended to other industries and domains that require secure and efficient identity management, such as supply chain management, healthcare, and IoT. This demonstrates the versatility and potential of the blockchain-based VDKMS approach.

\section{Conclusion and Future Work} \label{sec:conclusion}
We have presented a novel Decentralized Cellular Vehicular-to-Everything Key Management System (\systemname{}) that leverages advancements in Self-Sovereign Identity (SSI) and Decentralized Identifier (DID) technology to provide a secure, scalable, and efficient solution for vehicular communications. We have designed the system to address the unique challenges and requirements associated with vehicular networks, while also ensuring compatibility and interoperability with existing communication standards and protocols.
While our work has demonstrated the potential of the proposed approach, there are several research directions that we plan to explore in the future: (1) Integration of additional privacy-enhancing technologies, such as zero-knowledge proofs or secure multi-party computation, to further improve the privacy guarantees offered by the system. (2) Investigation of alternative consensus mechanisms that can provide greater scalability and energy efficiency while maintaining the required security and trust properties. (3) Extension of the proposed DKMS to support more advanced vehicular communication scenarios, such as platooning, cooperative maneuvering, and other collaborative driving applications. (4) Evaluation of the system's performance and applicability in large-scale, real-world deployment scenarios, involving a diverse range of vehicles and communication environments. Addressing these future research directions will further enhance the capabilities of the proposed \systemname{} and contribute to the ongoing development of secure, reliable, and efficient vehicular communication systems.

\section*{Acknowledgment}
We gratefully acknowledge this research is partially supported by FHWA EAR 693JJ320C000021.

\bibliographystyle{IEEEtran}
\bibliography{reference}
\end{document}